\documentclass[aps,prl,twocolumn,superscriptaddress,showpacs]{revtex4}

\usepackage{graphicx}

\begin{document}
\title{Impurity-induced transition to a Mott insulator in Sr$_3$Ru$_2$O$_7$}

\author{R. Mathieu\cite{cryo}}
\affiliation{Spin Superstructure Project (ERATO-SSS), JST, AIST Central 4, Tsukuba 305-8562, Japan}

\author{A. Asamitsu}
\affiliation{Spin Superstructure Project (ERATO-SSS), JST, AIST Central 4, Tsukuba 305-8562, Japan}
\affiliation{Cryogenic Research Center (CRC), University of Tokyo, Bunkyo-ku, Tokyo 113-0032, Japan}

\author{Y. Kaneko}
\affiliation{Spin Superstructure Project (ERATO-SSS), JST, AIST Central 4, Tsukuba 305-8562, Japan}

\author{J. P. He}
\affiliation{Spin Superstructure Project (ERATO-SSS), JST, AIST Central 4, Tsukuba 305-8562, Japan}

\author{X. Z. Yu}
\affiliation{Spin Superstructure Project (ERATO-SSS), JST, AIST Central 4, Tsukuba 305-8562, Japan}

\author{R. Kumai}
\affiliation{Correlated Electron Research Center (CERC), AIST Central 4, Tsukuba 305-8562, Japan}

\author{Y. Onose}
\affiliation{Spin Superstructure Project (ERATO-SSS), JST, AIST Central 4, Tsukuba 305-8562, Japan}

\author{N. Takeshita}
\affiliation{Correlated Electron Research Center (CERC), AIST Central 4, Tsukuba 305-8562, Japan}

\author{T. Arima}
\affiliation{Spin Superstructure Project (ERATO-SSS), JST, AIST Central 4, Tsukuba 305-8562, Japan}
\affiliation{Institute of Multidisciplinary Research for Advanced Materials, Tohoku University, Sendai 980-8577, Japan}

\author{H. Takagi}
\affiliation{Correlated Electron Research Center (CERC), AIST Central 4, Tsukuba 305-8562, Japan}
\affiliation{Department of Advanced Materials Science, University of Tokyo, Kashiwa 277-8581, Japan}
\affiliation{CREST, Japan Science and Technology Corporation (JST)}

\author{Y. Tokura}
\affiliation{Spin Superstructure Project (ERATO-SSS), JST, AIST  Central 4, Tsukuba 305-8562, Japan}
\affiliation{Correlated Electron Research Center (CERC), AIST Central 4, Tsukuba 305-8562, Japan}
\affiliation{Department of Applied Physics, University of Tokyo, Tokyo 113-8656, Japan}

\begin{abstract}

The electrical, magnetic, and structural properties of Sr$_3$(Ru$_{1-x}$Mn$_x$)$_2$O$_7$ (0 $\leq x \leq$ 0.2) are investigated. The parent compound Sr$_3$Ru$_2$O$_7$ is a paramagnetic metal, critically close to magnetic order. We have found that, with a Ru-site doping by only a few percent of Mn, the ground state is switched from a paramagnetic metal to an antiferromagnetic insulator. Optical conductivity measurements show the opening of a gap as large as 0.1 eV, indicating that the metal-to-insulator transition is driven by the electron correlation. The complex low-temperature antiferromagnetic spin arrangement, reminiscent of those observed in some nickelates and manganites, suggests a long range orbital order.

\end{abstract}

\date{\today}

\pacs{74.70.Pq, 71.27.+a, 71.30.+h}

\maketitle

The Ruddelson-Popper-type Sr$_{n+1}$Ru$_{n}$O$_{3n+1}$ series show metallic properties, as well as exotic superconductivity and spin/orbital order. SrRuO$_3$ ($n = \infty$, perovskite) and Sr$_4$Ru$_3$O$_{10}$  ($n$ = 3)  are itinerant ferromagnets with Curie temperatures ($T_c$) around 160 K and 100 K respectively\cite{SRO113,SRO4310}. In contrast, the single-layered Sr$_2$RuO$_4$ ($n$ = 1) and double-layered Sr$_3$Ru$_2$O$_7$ ($n$ = 2) do not show ferromagnetism (FM). Sr$_2$RuO$_4$ is a well known superconductor, which displays the unconventional spin-triplet pairing\cite{SRO214}. Sr$_3$Ru$_2$O$_7$ (Ru$^{4+}$, 4d$^4$) is essentially paramagnetic\cite{SRO327} although there is some controversy in the literature related to the presence of ferromagnetic SrRuO$_3$ and Sr$_4$Ru$_3$O$_{10}$ impurities. A metamagnetic transition is observed at low temperatures in large magnetic fields\cite{SRO327QCP,SRO327HC}, indicating the presence of quantum criticality. Another signature of this phase competition is found in hydrostatic pressure experiments\cite{SRO327press} or Sr-site doping by Ca\cite{SCRO327} of Sr$_3$Ru$_2$O$_7$,  which stabilizes the antiferromagnetic (AFM) state. In the case of Ca$_3$Ru$_2$O$_7$, which has a lattice structure closely related to that of Sr$_3$Ru$_2$O$_7$, the AFM ordering is observed at low temperature ($\sim$ 56 K). The system remains metallic below this temperature, but near 48 K, a first-order-like transition to a less conductive state is observed\cite{Yoshida}.

In this article, we investigate the effects of the Ru-site doping by Mn of Sr$_3$Ru$_2$O$_7$. A metal-to-insulator transition is observed as the Mn content increases. Optical conductivity measurements reveal the opening of a Mott-like gap\cite{Mott}. Neutron diffraction experiments show that the low temperature insulating state is associated with a novel antiferromagnetic spin arrangement, which suggests some complex orbital order. Interestingly, while there are many observations of insulator-to-metal transitions upon tuning of one-electron bandwidth or band-filling by doping\cite{Imada}, there are few reports of the opposite, i.e. of the formation of a Mott-like insulating state upon addition of impurities in a metal. One example of such impurity doping effects is found in the Mott insulator V$_2$O$_3$. As V$^{3+}$ is replaced by Ti$^{3+/4+}$, the system becomes metallic by modification of the band filling and/or bandwidth\cite{V2O3}. However, the replacement of V$^{3+}$ by Cr$^{3+}$ instead strengthens the electron correlation (via modification of the lattice and hence the bandwidth), so that the metal-to-insulator transition occurs at higher temperatures than in the undoped compound\cite{V2O3}. In the present case of Sr$_3$(Ru$_{1-x}$Mn$_x$)$_2$O$_7$, the effect of the impurity doping is even more dramatic. A few percents of Mn yield a drastic phase change, turning the paramagnetic metal Sr$_3$Ru$_2$O$_7$ ($x$ = 0) into an antiferromagnetic Mott insulator ($x$ $>$ 0.025).
 
Single crystals of Sr$_3$(Ru$_{1-x}$Mn$_x$)$_2$O$_7$ ($x$ = 0, 0.005, 0.025, 0.05, 0.075, 0.1, and 0.2) were grown by the floating zone method. The Mn composition $x$ was checked by inductively coupled plasma (ICP) analysis. X-ray diffraction and SQUID magnetometry reveal minor impurities of Sr$_2$RuO$_4$ and traces of the ferromagnetic Sr$_4$Ru$_3$O$_{10}$. The crystals easily cleave along shiny $ab$-planes. The in-plane resistivity $\rho$ of the crystals was measured as a function of the temperature $T$ using a standard four-probe method on a PPMS6000 system. The heat capacity was recorded using a relaxation method from $T$ = 0.6 K to 100 K with the same measurement system. X-ray and neutron diffraction data were collected at selected temperatures with setups using closed cycle helium refrigerators, for the crystal with $x$ = 0.05: The x-ray data was recorded on a Rigaku SPD curved imaging plate system by using Si(111)-double-crystal monochromated synchrotron radiation x-ray ($\lambda$= 0.688 {\AA}) at the beam line BL-1A of the Photon Factory, High-Energy Accelerator Research Organization (KEK), Japan. The cell parameters were calculated from the diffraction peaks of the x-ray oscillation photographs of the single crystal assuming a tetragonal ($I4/mmm$) symmetry in the whole temperature region. 

\begin{figure}[h]
\includegraphics[width=0.46\textwidth]{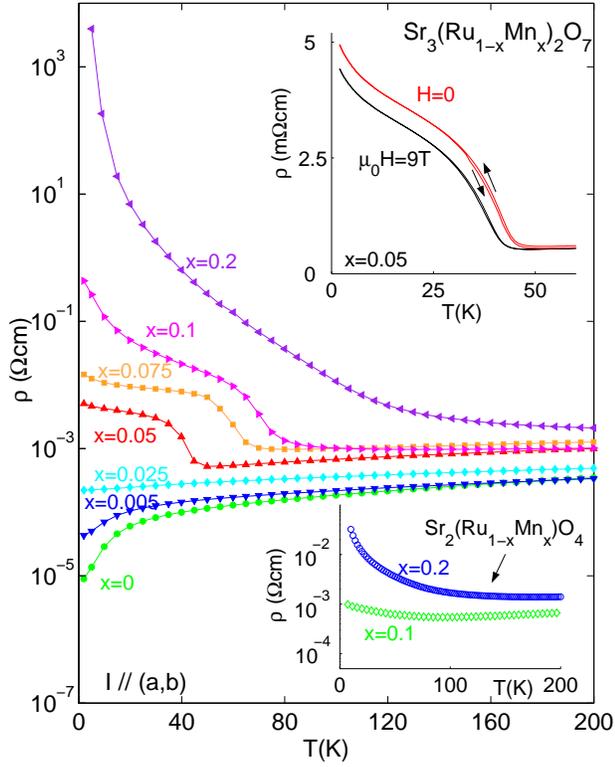}
\caption{(color online) Temperature $T$ dependence of the resistivity $\rho$ of Sr$_3$(Ru$_{1-x}$Mn$_x$)$_2$O$_7$  for different Mn doping $x$. The electrical current was fed in the $ab$-plane, and in the experiments shown in the upper inset, a magnetic field of $\mu_0H$ = 9T was applied along the $c$-axis of the crystal with $x$ = 0.05. The resistivity was recorded on cooling and heating, as indicated by the arrows. The lower inset shows the weaker effect of such a substitution in the case of single crystals of Sr$_2$(Ru$_{1-x}$Mn$_x$)O$_4$ (from Ref. {\protect \onlinecite{Hatsuda}}) for comparison; this effect is as well weak in the case of Sr(Ru$_{1-x}$Mn$_x$)O$_3$ (see Ref. {\protect \onlinecite{Miyasato}}) for similar amounts of Mn.}
\label{fig-resis}
\end{figure}

\noindent The powder neutron diffraction patterns were collected using a time-of-flight (TOF) diffractometer Vega, KEK, Japan.  The single crystal was pulverized for this purpose. The obtained powders were sealed with helium gas in a vanadium cell with a diameter of 9.2 mm.  TOF diffraction patterns for a wide $d$-range ( 1 - 20 {\AA}) were obtained at a 30-degree scattering bank.  The optical conductivity spectra $\sigma$($\omega$) of the crystal with $x$ = 0.1 were measured at different temperatures. Near-normal-incidence reflectivity spectra $R$($\omega$) were collected in the 0.05 - 3 eV energy range. Kramers-Kronig analysis was performed to derive $\sigma$($\omega$) from the measured $R$($\omega$).\\

\begin{figure}[h]
\includegraphics[width=0.48\textwidth]{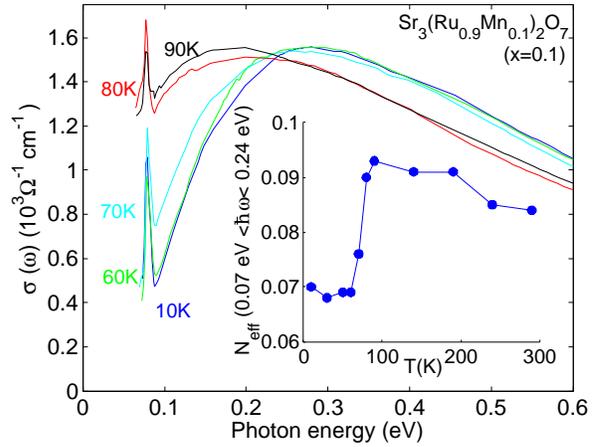}
\caption{(color online) Optical conductivity $\sigma$($\omega$) of Sr$_3$(Ru$_{0.9}$Mn$_{0.1}$)$_2$O$_7$ ($x$=0.1) at selected temperatures. The sharp peaks near 0.08 eV correspond to optical phonons coupled with the electronic excitations. The inset shows the low-energy spectral weight $N_{\rm eff}$ (effective number of electrons) plotted as a function of the temperature, obtained by integration of $\sigma$($\omega$) between 0.07 eV and 0.24 eV.}
\label{fig-opt}
\end{figure}

The dramatic effect of the Mn doping on the electrical properties of Sr$_3$(Ru$_{1-x}$Mn$_x$)$_2$O$_7$ is shown in Fig.~\ref{fig-resis}. While all the samples are metallic at room temperature, only the crystals with $x \leq$ 0.025 remain metallic at low temperatures. In the low-temperature metallic regime, the residual resistivity increases with $x$, indicating that Mn impurities act as scatterers. The crystals with $x >$ 0.025 in contrast show a sharp increase in resistivity at low temperatures. These resistivity jumps are accompanied by an insulating behavior at lower temperatures. The resistivity is discernibly affected by the application of a magnetic field, as seen in the upper inset for $x$ = 0.05. The resistivity increase due to the metal-insulator transition is shifted to lower temperature with an applied magnetic field. This suggests the antiferromagnetic nature of the lower-lying insulating state. 

The optical conductivity $\sigma$($\omega$) spectra are shown in Fig.~\ref{fig-opt} for the crystal with $x$ = 0.1 at different temperatures. Above $T$ = 80 K, the optical conductivity spectra show a broad peak in the mid-infrared region ($\sim$ 0.2 eV), which is typical of the incoherent metallic state\cite{Imada} in the vicinity of the Mott transition\cite{Mott}. While $\sigma$($\omega$) shows a minimal temperature variation above 80 K, the spectral weight in the low energy region is steeply suppressed below $T$ = 70 K, forming a gap-like structure. As a measure of the spectral weight, we utilize the effective number of electrons, defined as $N_{eff}(\omega)=(2m_0)/(\pi e^2N)\int_0^{\omega}\sigma(\omega')d\omega'$ \cite{Imada} where $m_0$ is the free electron mass and $N$ is the number of Ru ions per unit volume. The different $\sigma(\omega)$ curves shown in Fig.~\ref{fig-opt} define an isosbestic point (equal-absorption point, across which the spectral weights are transferred) near 0.24 eV, the typical energy scale of the Mott gap\cite{Imada}. Thus, by integrating $\sigma(\omega)$ in the 0.07 eV $<\hbar\omega<$ 0.24 eV energy range, we could estimate the low-energy spectral weight, as $N_{eff}$(0.07 eV $<\hbar\omega<$ 0.24 eV) = $N_{eff}$(0.24 eV) - $N_{eff}$(0.07 eV). As seen in the temperature variation of $N_{eff}$(0.07 eV $<\hbar\omega<$ 0.24 eV) shown in the inset of Fig.~\ref{fig-opt}, the formation of a gap occurs between $T$ = 60 and 80 K, in the vicinity of the resistivity-jump temperature (near $T$ = 78 K for the crystal with $x$ = 0.1, see Fig.\ref{fig-resis}). The Mn impurities increase the disorder. However, this large energy-scale ($\sim$ 0.2 eV) gap formation cannot be explained by disorder-induced localization effects\cite{Anderson}, as the typical energy scale of the Anderson localization is much smaller ($< 10^{-2}$eV). It is worth noting that such a large energy gap formation is observed in the optical spectra of Ca$_2$RuO$_4$\cite{Jung} and other Mott insulators\cite{Imada}. 

\begin{figure}[h]
\includegraphics[width=0.46\textwidth]{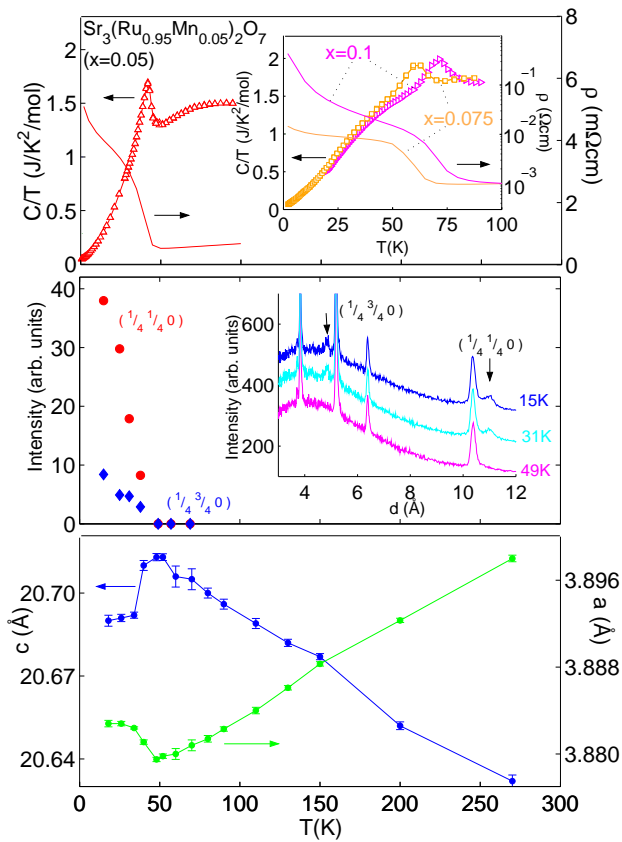}
\caption{(color online) Temperature dependence of different physical properties of Sr$_3$(Ru$_{0.95}$Mn$_{0.05}$)$_2$O$_7$ ($x$ = 0.05). Upper panel: heat capacity $C$ plotted as $C/T$ (left axis); the resistivity ($\rho$) data from Fig.~\ref{fig-resis} is added for comparison (right axis). The inset of this panel shows the corresponding data for the crystals with $x$ = 0.075 and $x$ = 0.1. Middle panel: intensity of the (1/4,1/4,0) and (1/4,3/4,0) magnetic peaks observed in neutrons diffractograms, as shown in the inset. Lower panel: tetragonal $a-$ and $c$-axis lattice parameters obtained from x-ray diffraction.}
\label{fig-all}
\end{figure}

We now investigate the magnetic and structural properties of the insulating state of the crystals with $x$ $\ge$ 0.05. We display in Fig.~\ref{fig-all} the temperature dependence of various physical properties of the system.  The low-temperature heat capacity $C$ data of the crystals with $x$ $\ge$ 0.05 (plotted as $C/T$) shown in the upper panel and inset of Fig.~\ref{fig-all} could be fitted with the contributions of conduction electrons (the $T$-linear coefficient $\gamma$ shows a weak sample dependence and amounts to 50 $\pm$ 5 mJ/mol/K$^2$) and phonon modes (the $T^3$-linear coefficient $\beta$ $\sim$ 1 $\pm$ 0.2 mJ/mol/K$^4$). The fit is slightly improved introducing a $T^2$-linear contribution, which may reflect a magnetic contribution to the heat capacity. More importantly, for the crystals with $x$ = 0.05 (main frame), and $x >$ 0.05 (inset), a peak is observed in the heat capacity in the vicinity of the resistivity jump temperature. In comparison, the heat capacity of the undoped compound (which remains a metal down to the lowest temperature) show no anomaly\cite{SRO327HC}. By integrating the data of the Mn doped crystals after subtraction of a certain baseline, an entropy change associated with the transition $\Delta S$ $\sim$ 0.45 $\pm$ 0.05 R could be estimated, implying the relevance of the spin entropy. This is confirmed by neutron diffraction experiments. As shown in the middle panel of Fig.~\ref{fig-all}, the magnetic peaks corresponding to AFM ordering show up below the metal-to-insulator transition temperature.

Furthermore, x-ray diffraction measurements show that around the same temperature the $c$-axis lattice parameter decreases, as seen in the lower panel of Fig.~\ref{fig-all} for $x$ = 0.05. This may indicate that the RuO$_6$ octahedra are compressed along the $c$-axis, in a Jahn-Teller like fashion, or that the buckling of the octahedra occurs. The electrical and magnetic properties of ruthenates are usually associated with the 4$d$ orbitals of Ru, and their hybridization with the $2p$ orbitals of oxygen. In Sr$_3$(Ru$_{1-x}$Mn$_x$)$_2$O$_7$, Ru$^{4+}$  is in the $S$ = 1 low spin state, so that the $e_g$ levels are empty, and only the $t_{2g}$ orbitals $d_{xy}$, $d_{yz}$, and $d_{zx}$ are partially filled. Structural modifications such as distortions of the RuO$_6$ octahedra yield changes in the relative energy of the $t_{2g}$ orbitals\cite{Fang}, and hence in the magnetic interaction. The high-temperature metallic state of Ca$_3$Ru$_2$O$_7$ was related to the predominant role of the $d_{xy}$ orbitals\cite{Yoshida}. The low-temperature insulating state was then associated to the increased contribution of the $d_{yz}$ and $d_{zx}$ orbitals promoted by lattice changes. However, the contraction of the $c$-axis observed by x-ray diffraction in Ca$_3$Ru$_2$O$_7$\cite{Ohmichi}, and in our Sr$_3$(Ru$_{1-x}$Mn$_x$)$_2$O$_7$ (c.f. Fig.~\ref{fig-all}) amounts merely to $\sim$ 0.1 \%, which is one order of magnitude lower than observed in systems with ferroic $d_{xy}$ orbital ordering such as Ca$_2$RuO$_4$\cite{Jung}.

\begin{figure}[h]
\includegraphics[width=0.46\textwidth]{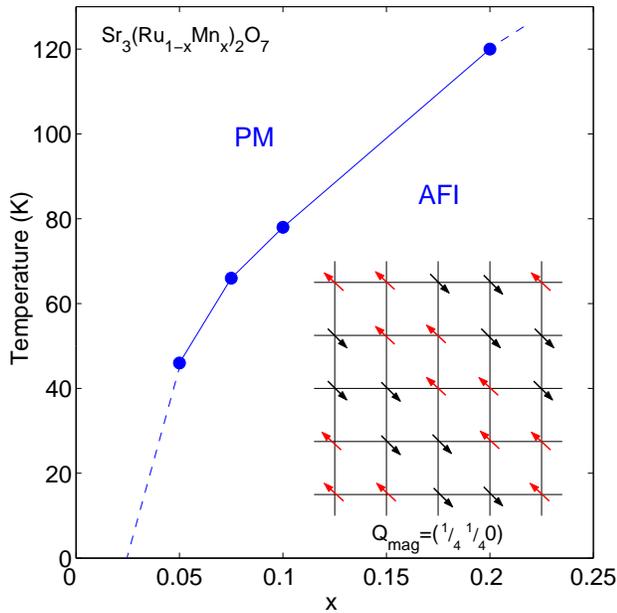}
\caption{(color online) Electronic phase diagram showing the phase boundary between the paramagnetic metal (PM) and antiferromagnetic insulator (AFI) regions. The boundary is plotted using transport data from Fig.~\ref{fig-resis}. A schematic view of the most probable magnetic arrangement in the $ab$-plane is shown in the inset using the neutron diffraction data collected for $x$ = 0.05 (another possible AFM structure is the same up-up-down-down arrangement with spins parallel to the $c$-axis).} 
\label{fig-phasediag}
\end{figure}

The measurement of the temperature dependence of various physical properties of Sr$_3$(Ru$_{1-x}$Mn$_x$)$_2$O$_7$ enables us to draw the schematic electronic phase diagram shown in Fig.~\ref{fig-phasediag}, which defines the phase boundary between the paramagnetic metal (PM) and the antiferromagnetic insulator (AFI). Antiferromagnetic peaks appear in neutron diffractograms at the metal-to-insulator transition temperatures (no additional peaks appear in the high-$Q$ region, nor ($n$/2 1/4 0) peaks). The most probable magnetic arrangement in the $ab$-plane is an up-up-down-down arrangement (with spins in the $ab$-plane or parallel to the $c$-axis) with the magnetic wave vector of (1/4 1/4 0), although this has to be confirmed by a thorough diffraction study. This spin arrangement is reminiscent of the zigzag chains of the E-type structure\cite{Hotta} observed at low temperatures in manganites with significant GdFeO$_3$-type distortion such as HoMnO$_3$\cite{HMO}, as well as in (Nd/Pr)NiO$_3$\cite{Garcia} nickelates. In $R$MnO$_3$ ($R$ being a rare earth ion), the amount of GdFeO$_3$-type distortion determines the strength of the superexchange interaction between Mn next-nearest-neighbors. For small $R$ ions, like Ho, the large superexchange interaction, coupled to the orbital order, yields the spin frustration\cite{HMO} and the appearance of the E-type antiferromagnetic state. Hence the complex magnetic structure of Sr$_3$(Ru$_{1-x}$Mn$_x$)$_2$O$_7$ with $x$ $>$ 0.025 may reflect some long-range order in the orbital sector\cite{Dagotto,Garcia}. The $c$-axis contraction observed in the vicinity of the resistivity jump is similar to that observed in Ca$_3$Ru$_2$O$_7$, as well as in Ca$_{2-x}$Sr$_x$RuO$_4$ (0.2 $< x<$ 0.5, so called ``Phase II'')\cite{Steffens} which displays a complex orbital order. However, the low temperature insulating AFM state of Sr$_3$(Ru$_{1-x}$Mn$_x$)$_2$O$_7$ ($x$ $\ge$ 0.05) does not result from such structural distortions as in the less conductive/two-dimensional metallic state of Ca$_3$Ru$_2$O$_7$ or Ca-rich (Sr,Ca)$_3$Ru$_2$O$_7$\cite{Ohmichi}, but from the local modification of the orbital states induced by the Mn impurities.

In summary, the physical properties of single crystals of the Sr$_3$(Ru$_{1-x}$Mn$_x$)$_2$O$_7$ (0 $\leq x \leq$ 0.2) system have been investigated. A metal-to-insulator transition, associated with the formation of a large charge gap ($\sim$ 0.1 eV), is observed as the Mn concentration $x$ increases. The closely correlated temperature-dependence of various physical quantities suggests a tight coupling of the lattice, orbital and spin degrees of freedom. Interestingly, the low-temperature insulating state shows a complex antiferromagnetic spin structure, which suggests some long-range order in $t_{2g}$ orbital sector. The results illustrate how the impurity doping (substituting only a few percents of Ru) allows the phase control of the multicritical layered ruthenates.

We thank Professors T. Ishigaki and T. Kamiyama for their help with the neutron diffraction measurements.

\end{document}